\documentclass[
preprint,
 amsmath,amssymb,
 aps,
]{revtex4-2}

\usepackage{graphicx}% Include figure files
\usepackage{dcolumn}% Align table columns on decimal point
\usepackage{bm}% bold math

\begin{document}

\title{Stark effect tunable terahertz transitions in finite carbon chains}
%\title{Tuning terahertz transitions in linear carbon chains}

\author{R. A. Ng}
\affiliation{
Physics Department, De La Salle University, 2401 Taft Avenue, 0922 Manila, Philippines
}

\author{M. E. Portnoi}
\affiliation{
Physics and Astronomy, University of Exeter, Stocker Road, Exeter EX4 4QL, United Kingdom
}

\author{R. R. Hartmann}
\email{richard.hartmann@dlsu.edu.ph}
\affiliation{
Physics Department, De La Salle University, 2401 Taft Avenue, 0922 Manila, Philippines
}

\begin{abstract}
We employ a tight-binding model to calculate the optical selection rules of gold-terminated carbyne chains in the presence of an applied electric field. We show that both the magnitude of the edge-state gap and the strength of optical transitions across it can be tuned via the Stark effect. In the case of sufficiently long carbyne chains, the dipole transitions between edge states occur within the THz frequency range.
\end{abstract}

\maketitle

%\section{\label{sec:introduction}Introduction}
The terahertz (THz) region of the electromagnetic spectrum lies between the microwave and infrared frequencies. Research interest in THz radiation is driven by its wide range of practical applications, ranging from medical imaging and materials characterization to wireless communication and data transfer. However, the many challenges associated with generating and manipulating THz waves have resulted in this frequency range often being referred to as the ``THz gap"~\cite{dragoman2004terahertz, lee2007searching}. One promising way to close the THz gap is to use low-dimensional carbon allotropes, such as graphene, carbon nanotubes, cyclocarbons~\cite{ng2022tuning} and carbynes as the building blocks of THz detectors and emitters~\cite{hartmann2014terahertz,hartmann2021terahertz}.

In contrast to graphene and carbon nanotubes, carbyne (a chain of carbon atoms with alternating single and triple bonds~\cite{casari2016carbon, zhang2020review}) exhibits strong photoluminescence~\cite{pan2015carbyne}, and its optical selection rules are governed by the number of atoms in the chain. There are a large variety of synthesis methods for producing carbyne~\cite{johnson1972silylation, gibtner2002end, chalifoux2010synthesis, nishide2006single, shi2016confined, cataldo2004synthesis, pan2015carbyne, kutrovskaya2020excitonic}, with recent advances in carbyne synthesis now allowing for the creation of thin films of highly aligned carbyne chains terminated by gold clusters~\cite{portnoi2023polarization}. In contrast to an infinite carbyne crystal which possesses a band gap in the vicinity of $2-2.3$ eV~\cite{al2014electronic}, gold-terminated carbon chains possess two topologically protected mid-gap energy states (edge states)~\cite{hartmann2021terahertz}. In the absence of doping, these states correspond to the highest occupied molecular orbital (HOMO) and lowest unoccupied molecular orbital (LUMO) levels. The HOMO$-$LUMO gap is dictated by the length of the chain, and for long enough chains this gap falls within the THz regime. However, for practical device applications it is desirable to be able to tune the size of the THz gap.  Indeed, one of the major challenges in the development of THz technology and science is the dearth of compact and efficient sources that can be readily tuned across the THz frequency range.

The energy spectrum of many molecules can be modified by the application of an external electric field, i.e., the Stark effect~\cite{stark1914beobachtungen}. In a finite straight carbon chain possessing an even number of dimers, the molecular orbital wavefunctions are either even or odd~\cite{hartmann2021terahertz}. Therefore, perturbation theory results in the lowest-order correction to the energy levels being dependent on the square of the electric field strength. For the bulk states, at experimentally achievable electric fields, this perturbation is negligibly small, far from reaching the Stark ladder regime~\cite{wannier1960wave}. However, for long enough chains, the perturbation to the edge state eigenvalues can be of the same order as the eigenvalues themselves. In this regime, the shift in the edge state eigenvalues due to the Stark effect cannot be described by second-order perturbation theory. 

\begin{figure}
    \includegraphics[width=0.35\textwidth]{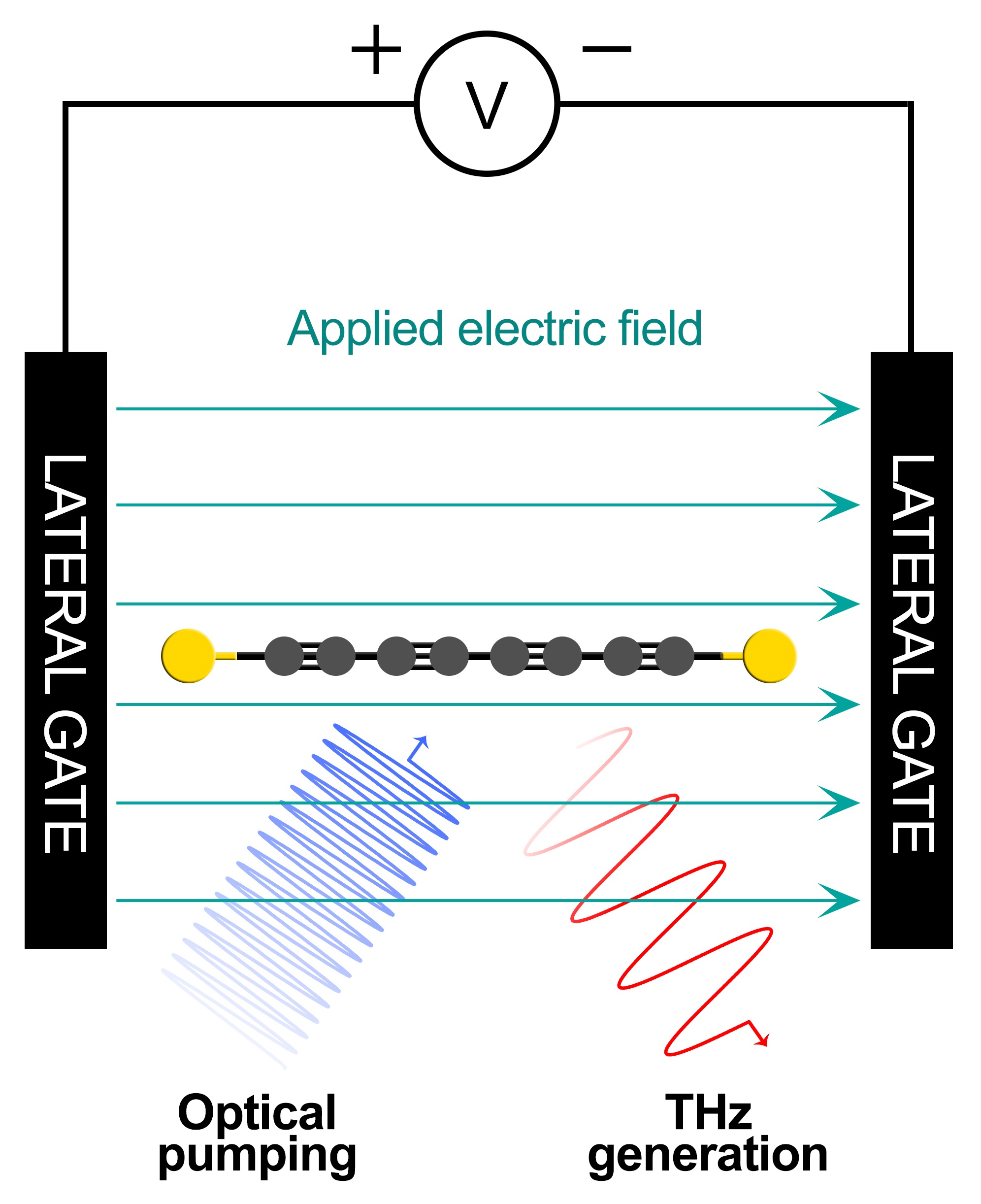}
    \caption{A schematic representation of carbyne with a polyynic structure (i.e., alternating single and triple bonds) terminated by gold nanoparticles. The chain is subjected to an electric field (depicted by the green arrows), $\boldsymbol{E}$, directed along the chain.}
    \label{fig:exp_setup}
\end{figure}

In what follows, we utilize the nearest-neighbor tight-binding approximation~\cite{dresselhaus1998physical} to compute the complete energy spectrum of gold-terminated carbyne chains subjected to an electric field aligned along the chain. We then derive an analytic expression for the edge-state gap for a chain composed of $N$ atoms under the influence of an electric field, as well as numerically calculate the oscillator strength of transitions involving these edge states. We demonstrate that the proposed THz generation scheme in Ref.~\cite{hartmann2021terahertz}, which was previously limited to emitting a single THz frequency, can be made tunable by incorporating an electric field. We show that when the gold atoms have the same on-site energies, increasing the electric field enlarges the edge-state gap but weakens transitions across it. Conversely, when the gold atoms possess differing on-site energies, increasing a subcritical electric field can reduce the edge-state gap size while enhancing the oscillator strength across the gap.

%\section{Theory}
We will now introduce a simple and tractable model that can be employed to describe the optical transitions among the electronic states of a finite-length carbyne chain terminated by gold nanoparticles, subjected to an applied electric field directed along the chain (see Fig.~\ref{fig:exp_setup}). Within the framework of the nearest neighbor tight-binding approximation~\cite{dresselhaus1998physical}, which has been successfully employed to describe the optical and electronic properties of various low-dimensional allotropic forms of carbon, the effective matrix Hamiltonian of a gold-terminated carbyne chain can be expressed as:~\cite{hartmann2021terahertz}
\begin{equation}
\mathbf{H}=\left(
\begin{array}{ccccc}
\Delta_{1} & \beta\\
\beta & 0 & \alpha\\
 & \alpha & \ddots & \ddots\\
 &  & \ddots & \ddots & \beta\\
 &  &  & \beta & \Delta_{2}
\end{array}
\right),
\label{eq:Ham_0}
\end{equation}
where $H_{j,j+1}=H_{j+1,j}=\beta$ if $j$ is odd, and $H_{j,j+1}=H_{j+1,j}=\alpha$ if $j$ is even. Here $\alpha$ and $\beta$ are the transfer integrals associated with the short (triple) and long (single) covalent bonds respectively, and $\Delta_{1}$ and $\Delta_{2}$ are the on-site energies of the edge atoms measured with respect to the on-site energies of the carbon atoms.

The eigenfunctions of the Hamiltonian Eq.~(\ref{eq:Ham_0}), which can be found in Ref.~\cite{kouachi2006eigenvalues,hartmann2021terahertz}, can be divided into two categories: the eigenfunctions corresponding to edge states and those corresponding to bulk states.  There are two edge states, and for the case of $\Delta_{1}=\Delta_{2}$ one is even, the other odd in parity. The remaining eigenfunctions, are bulk states, and for the case of $\Delta_{1}=\Delta_{2}$, are ordered in alternating parity with increasing energy. It should be noted that for long enough chains, i.e., when $N^2\gg1$, %where $N$ is the number of atoms in the chain, 
the eigenvalues can be expressed as an analytic approximation~\cite{hartmann2021terahertz}.

%The band gap of an infinite carbyne crystal depends only on the difference of the tight-binding hopping parameters. For a chain comprised of an odd number of atoms, $E=0$ is always a solution of the secular equation. And for an even number of atoms, there are two edge states. 

In the presence of an applied electric field, $\boldsymbol{E}$, the perturbation to the Hamiltonian is given by the expression
\begin{equation}
   \delta H_{i,j} = eEa_{\mathrm{CC}}\left(-\frac{N+1}{2}+j\right) \delta_{ij},
\end{equation}
where $e$ is the elementary charge. In this model, we employ the approximation that the interatomic distance ($a_{\mathrm{CC}}$) is uniform throughout the chain. Typically, the characteristic polynomial of $\mathrm{det}\left(\textbf{H}+\delta\textbf{H}-\varepsilon\textbf{I}\right)=0$ needs to be solved numerically. Nevertheless, under the condition of $N^2\gg1$, the eigenvalues of the edge states can be approximated analytically. Since $\alpha^2,\beta^2\gg \varepsilon^2,a_{\mathrm{CC}}^2,\Delta_{1,2}^2$, the characteristic equation, $\mathrm{det}\left(\textbf{H}+\delta\textbf{H}-\varepsilon\textbf{I}\right)=0$, can be approximated as a quadratic in terms of $\varepsilon$. Solving this quadratic equation in the long-chain limit, the edge state %HOMO and LUMO 
eigenvalues can be approximately expressed by
\begin{equation}
    \varepsilon_{\pm}=\frac{\alpha^{2}-\beta^{2}}{\alpha^{2}}\left[\frac{1}{2}\left(\Delta_{1}+\Delta_{2}\right)\pm\sqrt{\left(V_{0}-\frac{\Delta_{1}-\Delta_{2}}{2}\right)^{2}+\alpha^{2}\left(\frac{\beta}{\alpha}\right)^{N}}\right],
\label{eq:edge_eigen}
\end{equation}
where $V_{0}=\frac{1}{2}eEa_{\mathrm{CC}}\alpha^{2}\left[\left(N-1\right)\alpha^{2}-\left(N+3\right)\beta^{2}\right]/\left(\alpha^{2}-\beta^{2}\right)^{2}$ and the $+$ and $-$ signs correspond to the LUMO and HOMO level of an undoped system, respectively. It should be noted that while perturbation theory is a valid approach for calculating corrections to bulk state eigenvalues, the first-order perturbation correction is zero. Consequently, the perturbation caused by an applied electric field on the bulk states is negligible.

\begin{figure}
\includegraphics[width=0.4\textwidth]{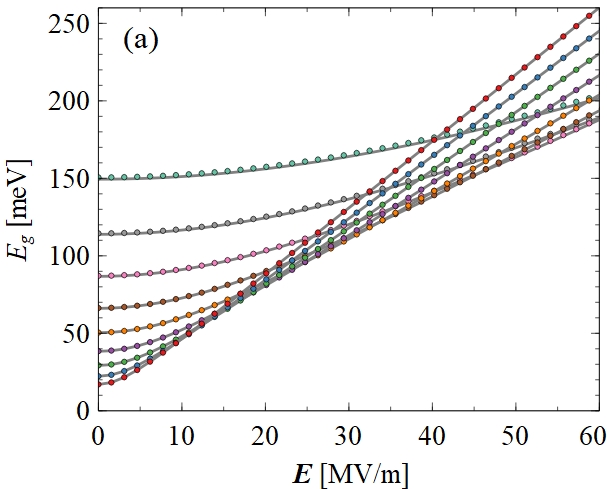}
\includegraphics[width=0.4\textwidth]{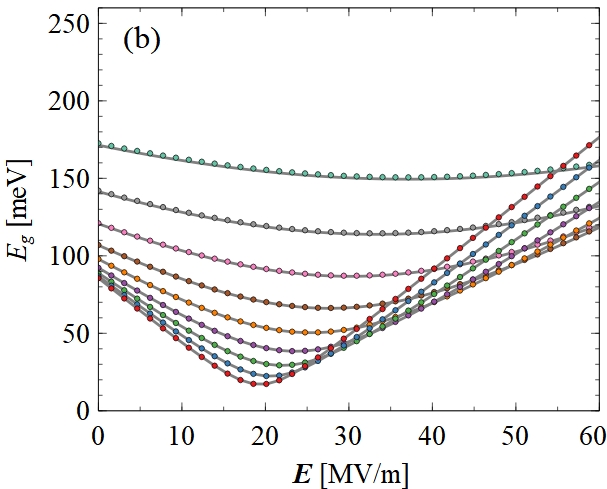}
    \caption{The edge-state gap dependence on the applied electric field strength for (a) the same on-site edge atom energies, $\Delta_1=\Delta_2=0.1$~eV, and (b) asymmetric on-site edge atom energies, $\Delta_1=0.2$~eV and $\Delta_2=0$. From top to bottom, the curves are arranged in increasing order of length, with the top-most curve corresponding to a chain of 24 atoms and the bottom-most curve 40 atoms. The dots represent the numerically determined edge-state gap, while the gray curves are obtained from the approximate eigenvalue expressions.
}
    \label{fig:HL_Gap}
\end{figure}

It can be seen from Eq.~(\ref{eq:edge_eigen}) that the edge-state-induced gap is highly tunable through the application of an electric field. For example, utilizing the tight-binding parameters of Ref.~\cite{al2014electronic} and $\Delta_1=\Delta_2=0.1$, a chain consisting of 36 atoms would result in an %HOMO$-$LUMO 
edge-state gap of 7.1~THz. Application of an electric field of strength $10$~MV/m increases the gap by 64.1\% to 11.6~THz. In the case of sufficiently long chains, the edge-state gap exhibits linear growth in response to an increasing electric field strength. In Fig.~\ref{fig:HL_Gap}~(a) we plot the dependence of the edge-state gap, $E_g$, on the applied electric field, for chains of various lengths, using the tight-binding parameters $\alpha=-4.657$~eV, $\beta=-3.548$~eV~\cite{al2014electronic}, with $\Delta_1=\Delta_2=0.1$~eV. The gray lines depict the edge-state gap obtained from the approximate expression, Eq.~(\ref{eq:edge_eigen}), whereas the dots represent the numerically determined edge-state gap.

Within the employed approximation and for sufficiently long chains, when $\Delta_{1}=\Delta_{2}$, the edge-state gap becomes independent of the magnitude of the edge atom's on-site energies, and it monotonically increases with higher field strengths. However, it should be noted that in the presence of edge atom on-site energy asymmetry, i.e., $\Delta_{1}\neq\Delta_{2}$, the system is analogous to a bipolar potential~\cite{hartmann2020bipolar} created by delta function potentials of differing signs located at each edge atom. Therefore, any asymmetry in edge atom on-site energies is equivalent to an effective internal electric field. Thus when the external field is anti-parallel to the internal field, i.e., $\Delta_{1}>\Delta_{2}$, and the applied electric field is below a critical threshold, it can be employed to reduce the size of the edge state gap (see Fig.~\ref{fig:HL_Gap}(b)). However, beyond this critical field strength, the size of the gap will increase. But when the internal field is aligned with the external field, i.e., $\Delta_{1}<\Delta_{2}$, the applied electric field can only increase the size of the gap.

%However, it can be seen from Eq.(\ref{eq:edge_eigen}) that in stark contrast to the chains with the same  on-site energy of edge atoms, chains which have higher  on-site energies for the edge atoms on the negative potential side, i.e., $\Delta_{1}>\Delta_{2}$, experience drastically different behavior. In this instance, when the applied electric field is below a critical threshold, it can be employed to reduce the size of the edge state gap (see Fig. \ref{fig:HL_Gap}(c)). However, beyond this critical field strength, the size of the gap will increase. Furthermore, it is evident from Eq. (\ref{eq:edge_eigen}) that for $\Delta_{1}<\Delta_{2}$, the asymmetry in  on-site energy leads to an effective enhancement of the applied electric field strength.

\begin{figure}
\includegraphics[width=0.4\textwidth]{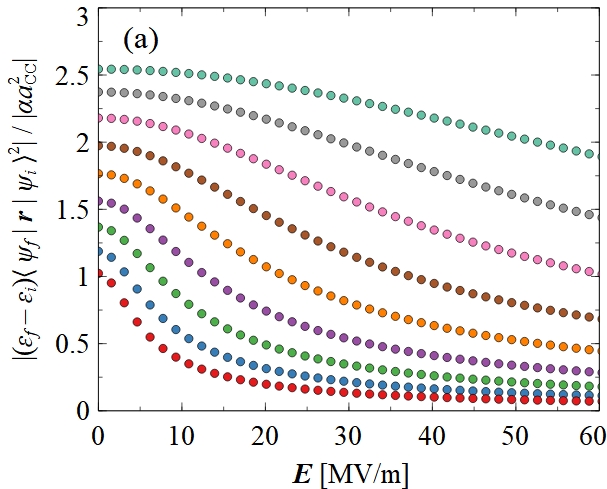}
\includegraphics[width=0.4\textwidth]{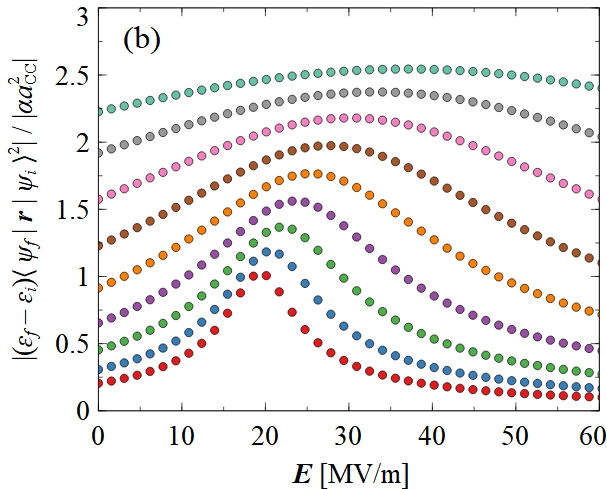}
    \caption{The dependence of the absolute value of the oscillator strength (normalized by a dimensionless constant) for the transition between edge states on the applied electric field strength, for a chain with: (a) equal on-site edge atom energies, $\Delta_1=\Delta_2=0.1$~eV; (b) different on-site edge atom energies, $\Delta_1=0.2$~eV and $\Delta_2=0$. From top to bottom, the curves are arranged in increasing order of length, with the topmost curve corresponding to a chain of 24 atoms and the bottommost curve 40 atoms. 
%(c) the HOMO$-$LUMO +1 transition for $\Delta_1=\Delta_2=0.1$ and (d) the the HOMO$-$LUMO +2 transition for $\Delta_1=\Delta_2=0.1$.
}
    \label{fig:osc}
\end{figure}
\begin{figure}
\includegraphics[width=0.4\textwidth]{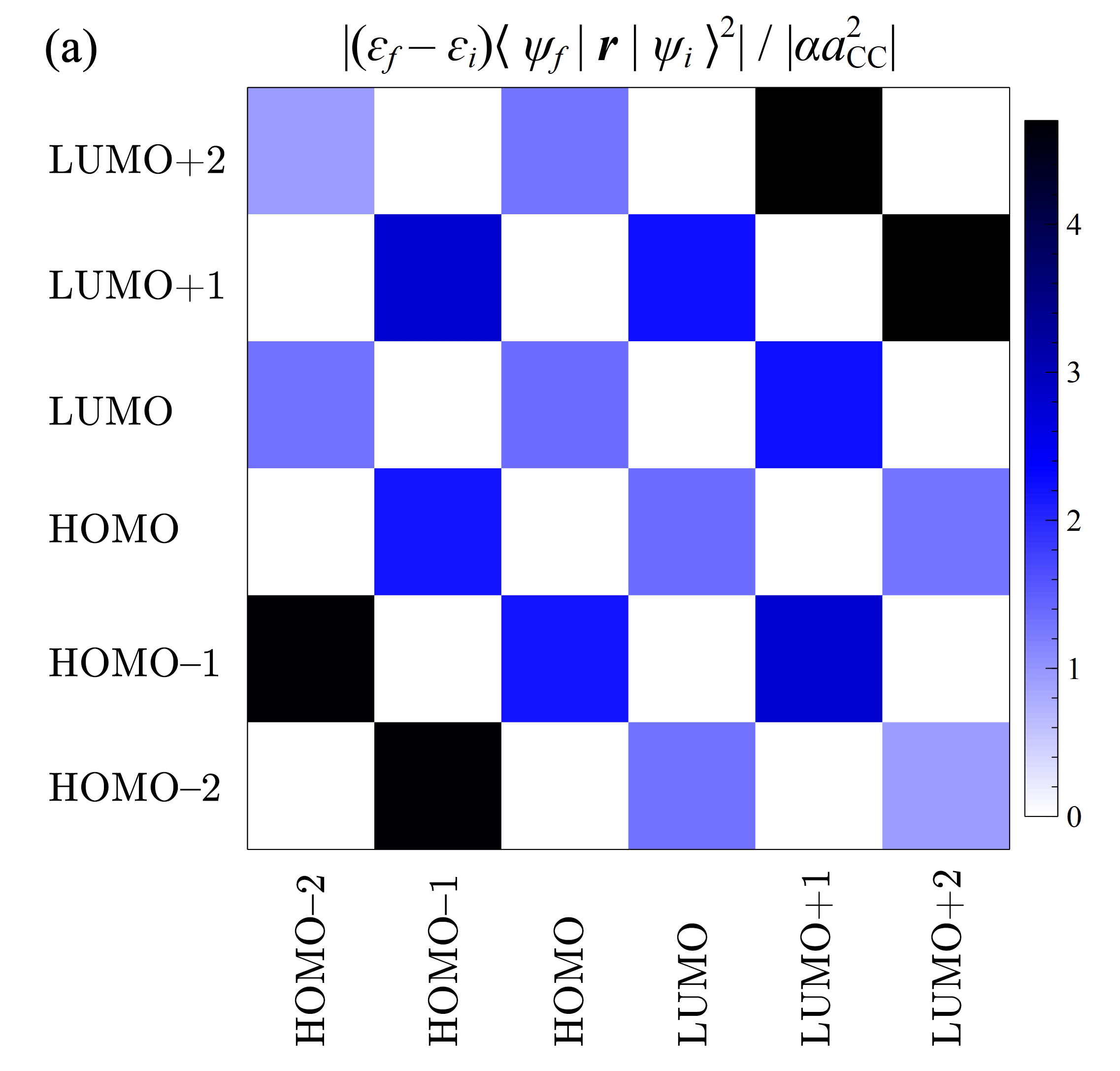}
\includegraphics[width=0.4\textwidth]{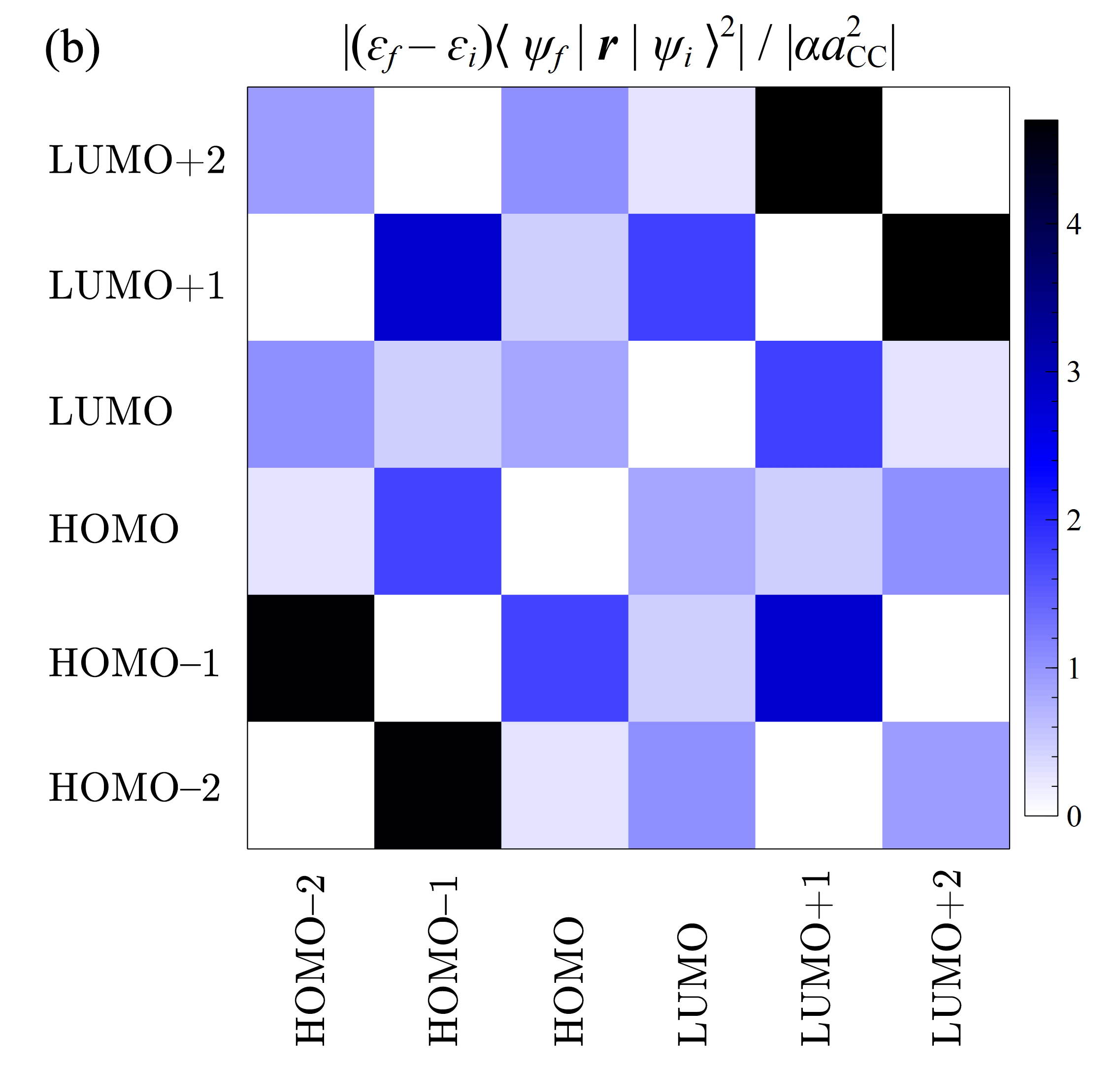}
\caption{The absolute value of the oscillator strength, normalized by a dimensionless constant, for inter-level transitions in a chain composed of $N = 36$ atoms, for electric field strengths: (a) $E = 0$ and (b) $E = 10$~MV/m. The tight-binding parameters are $\alpha = -4.657$ eV and $\beta = -3.548$~eV, while the on-site energies of the end atoms are $\Delta_1 = \Delta_2 = 0.1$~eV. The application of an electric field results in transitions that were previously forbidden becoming allowed, while strongly allowed transitions remain permissible.
}
\label{fig:osc_matrix}
\end{figure}

%\section{Optical transitions}
For light linearly polarized along the carbyne chain axis, the oscillator strength of a transition from initial state $i$ to final state $f$ may be defined as: 
\begin{equation}
   F_{if}=\frac{2}{3}\frac{m_{e}}{\hbar^{2}}\left(\varepsilon_{f}-\varepsilon_{i}\right)\left|\left\langle \psi_{f}\left|\boldsymbol{r}\right|\psi_{i}\right\rangle \right|^{2},
    \label{eq:oscillatorstrength}
\end{equation}
where $\psi_{i}$ and $\psi_{f}$ are the wave functions of the initial and final states, $\varepsilon_i$ and $\varepsilon_f$ are their corresponding energies, $m_e$ is the mass of an electron, %$\hat{\boldsymbol{e}}$ is the unit polarization vector of light, 
and $\boldsymbol{r}$ is the position operator.  

When the on-site energies of the edge atoms are identical, i.e., $\Delta_1=\Delta_2$, the oscillator strength of a transition between edge states exhibits a monotonic decrease as the applied electric field strength increases (see Fig.~\ref{fig:osc}~(a)). In longer chains, the oscillator strength exhibits a more pronounced decline with increasing electric field strength. Thus, there is a trade-off between increasing the gap size, and decreasing the oscillator strength.

For the case of $\Delta_1\neq\Delta_2$, the internal electric field arising from edge atom on-site energy asymmetry shifts the peak oscillator strength in the electric field. When the internal and external fields are opposed, i.e., the edge atom with the highest on-site energy is at the end at the lowest potential, the peak in oscillator strength occurs at a higher applied field (see Fig.~\ref{fig:osc}~(b)). In this instance, below a critical field strength, a decrease in the edge-state gap can coincide with an increase in the oscillator strength. 

The symmetry of the wave functions of the initial and final states strongly governs the optical selection rules. When the edge atom on-site energies are identical, i.e., $\Delta_1=\Delta_2$, under zero electric field, the parity of the wave functions for all states are purely even or odd, and the parity of consecutive energy levels alternate. Since the position operator appearing in Eq.~(\ref{eq:oscillatorstrength}) is an odd operator, optical transitions are allowed only when the parities of the initial and final states differ (see Fig.~\ref{fig:osc_matrix}~(a)). As mentioned previously, for experimentally achievable electric fields, the perturbation to the energy levels and wavefunctions of bulk states is negligible. Consequently, the oscillator strength of transitions between bulk states is not significantly affected by an external electric field. However, the electric-field-induced perturbation to the edge state eigenvalues is comparable in magnitude to the unperturbed eigenvalues themselves, and the eigenvectors undergo significant modification. 

\begin{figure}
\centering
\includegraphics[width=0.4\textwidth]{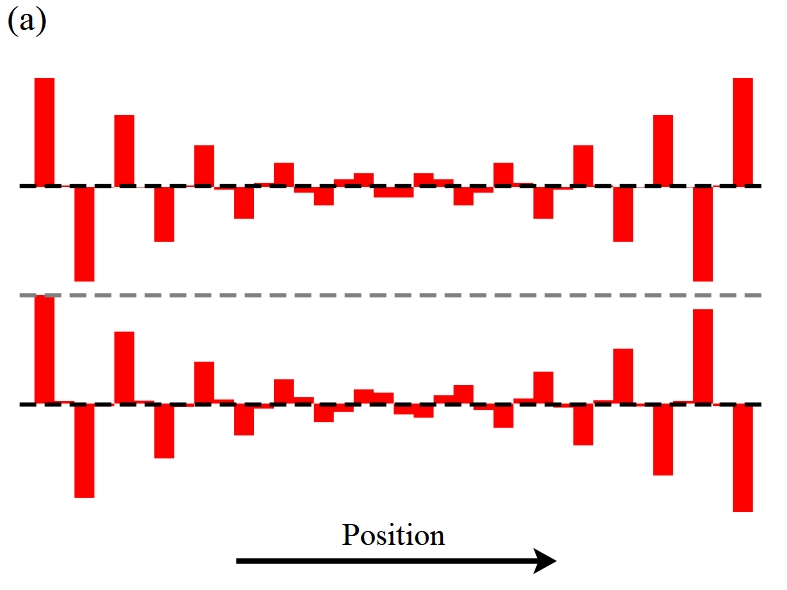}
\includegraphics[width=0.4\textwidth]{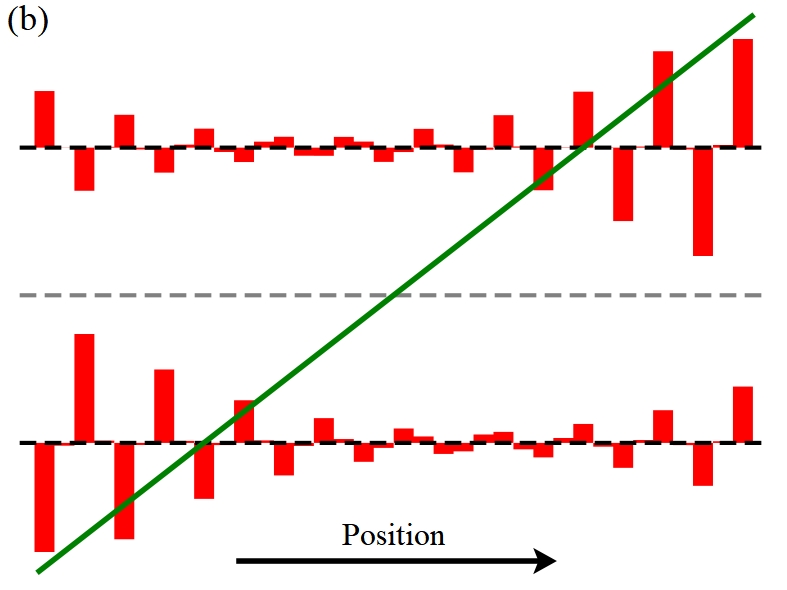}
\caption{A schematic diagram of the edge-state energy levels and wave functions of gold-terminated carbyne in (a) the absence and (b) presence of an applied electric field, of strength 10 MV/m.  The mid-gap energy of the undoped system is represented by the grey dashed line, while the positive and negative edge state energies are depicted by the black dashed lines, and the green line corresponds to the electrostatic potential. The displayed wave functions correspond to a carbyne chain composed of 36 atoms, characterized by the tight-binding parameters: $\alpha=-4.657$~eV, $\beta=-3.548$~eV, and $\Delta_1=\Delta_2=0.1$~eV.}
\label{fig:edge_well_barrier}
\end{figure}

When the edge atom on-site energies are the same, an applied electric field breaks the even-odd symmetry of the edge state eigenfunctions. As the electric field strength increases, the edge states becomes increasingly localized on one side of the chain compared to the other (see Fig.~\ref{fig:edge_well_barrier}). With each edge state becoming more localized on opposite sides of the chain. Thus for the inter-edge-state transition, the overlap of the edge state wave functions appearing in Eq.~(\ref{eq:oscillatorstrength}), reduces, leading to a suppression of the oscillator strength. Another consequence of each edge state becoming more localized to differing edges of the chain is that transitions between an edge state to a bulk states, which were previously forbidden due to symmetry, become stronger with increasing electric field strength, e.g., HOMO to LUMO+1 (see Fig.~\ref{fig:osc_matrix}). Conversely, transitions that were previously allowed become weaker as the electric field strength increases, e.g. HOMO to LUMO+2 (see Fig.~\ref{fig:osc_matrix}). 

As discussed previously, the presence of edge-atom on-site energy asymmetry is simply analogous to an effective internal electric field. This intrinsic field leads to the edge-state eigenfunctions being neither even nor odd. 
When an external field is applied anti-parallel to the internal field and reaches a critical strength, the fields are canceled out, and the edge state wavefunctions attain parity symmetry, resulting in a peak in the oscillator strength. Further increasing the electric field will break this symmetry, leading to a decrease in oscillator strength.

\begin{figure}
\includegraphics[width=0.5\textwidth]{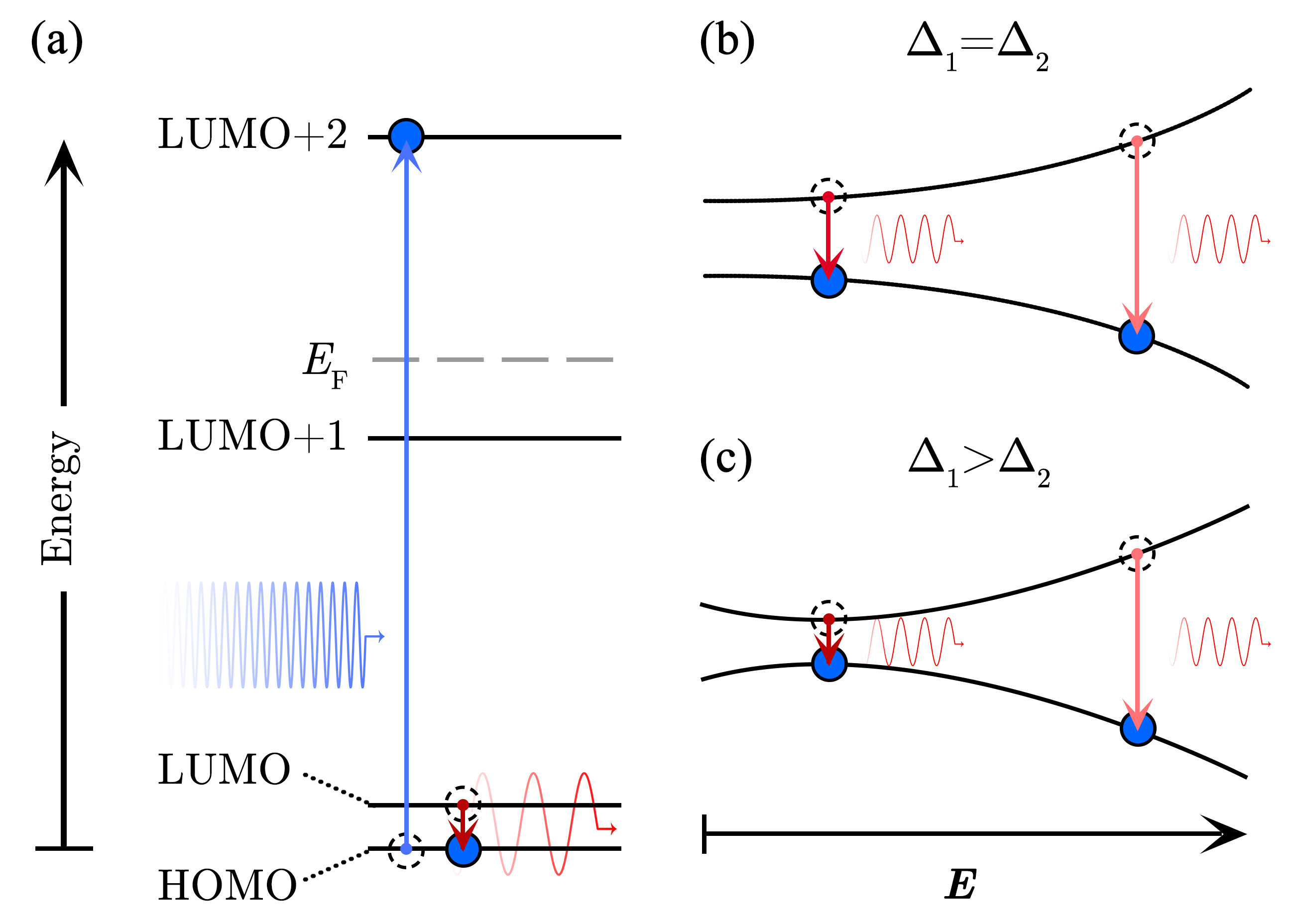}
\caption{
A depiction of a THz generation process within an n-doped carbyne chain terminated with gold nanoparticles, subject to an applied electric field. Panel (a) shows a high-frequency optical photon exciting an electron from the lower energy edge state to the energy level corresponding to the LUMO+2 of the undoped system, leaving behind an unoccupied state. Subsequently, an electron from the higher energy edge state can relax to fill this vacant state via the emission of a THz photon. Panels (b) and (c) show the dependence of the edge-state gap on the applied electric field, for the case of the edge atoms having the same and different on-site energies, respectively. The frequency of the emitted THz photon can be adjusted by varying the strength of the applied electric field.
}
\label{fig:THz_scheme}
\end{figure}

The high tunability of the optically active edge-state gap, coupled with robust edge-to-bulk state transitions, presents promising prospects for THz applications. Arrays of gold-terminated carbyne have the potential to generate THz radiation whose frequency can be selected by the applied device voltage via optical pumping (see Fig.~\ref{fig:THz_scheme}). Current synthesis methods often lead to the n-doping of carbyne chains by the gold nanoparticles~\cite{portnoi2023polarization}. In this instance, a high-frequency optical photon could promote an electron from the lower energy edge state to the level corresponding to the LUMO+2  of the undoped system. Subsequently, an electron can relax from the higher energy edge state to the lower energy edge state, emitting a THz photon in the process. The frequency of this emitted THz photon can be tuned by adjusting the strength of the applied electric field.

%\section{Conclusion}

In conclusion, we have shown that the size of the edge-state gap of finite-length carbyne chains, terminated by gold nanoparticles, can be tuned by an applied electric field aligned along the chain’s axis. The gap size response to the applied field is highly dependent on the relative on-site energies of the edge atoms. When the edge atoms possess the same on-site energies, then an applied field increases the size of the edge-state gap, while the oscillator strength decreases for transitions across it. However, when the edge atoms have non-equivalent on-site energies, an effective, internal electric field is established across the chain. The application of an anti-parallel external field can be employed to reduce the edge-state gap and enhance the oscillator strength of transitions across it. Whereas, the application of a parallel external field will increase the edge-state gap while reducing the oscillator strength. 

For sufficiently long chains, the edge-state gap falls within the THz range. The ability to tune the THz energy gap between optically active states positions gold-terminated carbyne as an excellent candidate for serving as the building block of a new class of tunable THz lasers. Furthermore, while our focus has been on carbon chains terminated by gold, the key findings of our study can be extended to a diverse range of systems that feature topologically protected states separated by a finite distance.

\begin{acknowledgments}
This work was supported by the EU H2020 RISE projects TERASSE (H2020-823878) and DiSeTCom (H2020-823728). R.A.N. is supported by the DOST-SEI ASTHRDP program. R.R.H. acknowledges financial support from URCO (15 F 2TAY21 - 3TAY22). M.E.P. was supported by NATO Science for Peace and Security Project No. NATO.SPS.MYP.G5860
\end{acknowledgments}

\bibliography{ref}

\end{document}